# Modelling of COVID-19 Using Fractional Differential Equations

Rishi Patel[a5], P. Sainani[a5], M. Brar[a], R. Patel[a5], X. Li[a,a1], J. Drozd[a3,a7], F. A. Chishtie[a2], A. Benterki[a6], T. C. Scott[a4], S. R. Valluri[a,b1]

[a]*Department of Physics and Astronomy, University of Western Ontario, 1151 Richmond Street, London Ontario, N6A 3K7, Canada*
[a]*Mathematics, King's University College, Western University, 266 Epworth Avenue, London Ontario, N6A 2M3, Canada,*
[a]*Department of Mechanical Engineering, Western University,*
[a]*Department of Occupational Science and Occupational Therapy, University of British Columbia, Vancouver, BC V6T 2B5*
[a]*Mathematics Department, Huron University, London Ontario, N6G 1H3,*
[a]*Institut für Physikalische Chemie, RWTH-Aachen University, 52056 Aachen, Germany,*
[a]*Department of Medical Sciences, University of Western Ontario,*
[a]*LMP2M Laboratory, Department of Mathematics, University of Médéa, Algeria,*
[a]*Lawrence Kinlin School of Business, Fanshawe College, London Ontario, N5Y 5R6*

## Abstract

In this work, we have described the mathematical modeling of COVID-19 transmission using fractional differential equations. The mathematical modeling of infectious disease goes back to the 1760s when the famous mathematician Daniel Bernoulli used an elementary version of compartmental modeling to find the effectiveness of deliberate smallpox inoculation on life expectancy. We have used the well-known SIR (Susceptible, Infected and Recovered) model of Kermack & McKendrick to extend the analysis further by including exposure, quarantining, insusceptibility and deaths in a SEIQRDP model. Further, we have generalized this model by using the solutions of Fractional Differential Equations to test the accuracy and validity of the mathematical modeling techniques against Canadian COVID-19 trends and spread of real-world disease. Our work also emphasizes the importance of Personal Protection Equipment (PPE) and impact of social distancing on controlling the spread of COVID-19.

## Introduction

The Mathematical Modelling of infectious diseases is a tool that can be used to study the spread and predict the course of outbreaks. One of its earliest accounts can be traced back to the 1760's when Daniel Bernoulli used a rudimentary version of compartmental modelling to determine the effectiveness of the deliberate smallpox inoculation on life expectancy [1]. In 1927, William O. Kermack and A.G. McKendrick pioneered the way for mathematical epidemiology. In their paper *"A Contribution to the Mathematical Theory of Epidemics"*, they established the fundamental SIR model where S, I, and R stand for the Susceptible, Infected, and Removed. From this model countless other epidemiological models have been derived throughout history [2].

Deadly infectious diseases with pandemic potential have plagued mankind for several thousands of years causing an innumerable death toll. Most recently, the emergence and rapid spread of severe acute respiratory syndrome coronavirus (SARS-CoV-2) which has to date led to more than 6 million deaths [3] and has led to a global pandemic. Although many Non-Pharmaceutical Intervention (NPI) strategies were implemented and successfully reduced the death toll and infection rate, their delayed implementation was unable to limit COVID-19. The efficiency of several preventative strategies and modelling the spread of the coronavirus outbreak are of great interest to society and public health [4], [5].

Through mathematical modelling, epidemiologists can gain a better understanding of the course of the disease and can use current data for predicting future trends to mitigate the risk of widespread toll [2]. Using epidemic curves, researchers can extrapolate disease data and trends to prepare for potential disease burden (economic/social) and monitor mutations that give rise to new variants. These models provide epidemiologists a mechanism for predicting and tracking the course of an outbreak–allowing better preparations for such afflictions.

Mathematical modelling is a well-known method that has been used for studying the recent COVID-19 outbreak. Using analytical methods, computer algebra software, fractional calculus (i.e., FDE-fractional differential equations), and the Lambert W Function, we propose a framework to study solutions of the SIR, SEIRS (Exposed[E]), and the fractional SEIQRDP (Quarantined[Q], Death[D], Insusceptible[P], Recovered[R]) models. These solutions will test the accuracy and validity of mathematical modelling techniques against Canadian COVID-19 trends and the spread of real-world disease. This paper also highlights the implications of personal protection equipment (PPE) and the impact of social distancing on managing the spread of COVID-19 [2], [6], [7].

The unstable nature of the COVID-19 virus and its high mutation rate led to the presence of several variants. Variants can be classified as variants being monitored, variants of interest, variants of concern, and variants of high consequence. The spread of symptoms of variants can differ depending on the effect and type of virus mutation that has occurred. When assessing variants, researchers observe the transmissibility, virulence, vaccine effectiveness, and diagnostic testing to decide which classification should be assigned to that variant. PPE also plays an integral role in the prevention of contracting COVID-19, but there are certain changes that manufacturers must make to their products. Variants tend to have changes in their transmissibility meaning manufacturers may have to create certain products that are better suited for protection in individuals that are at greater risk of contracting COVID-19 [8].

The SIR model compartmentalised the sample population into three parameters respectively: Susceptible, Infected and Removed (recovered and deceased) individuals. The dynamic nature of this model allows a simplified representation due to the nonlinear behaviour of transmission. The Kermack-McKendrick model's initial parameters can be modified as populations evolve to create even more specific models, to include parameters such as disease-acquired immunity, vaccinations, social interventions, among others. These modifications have been used to model outbreaks such as measles, rubella, HIV/AIDS and most recently, COVID-19 [2], [9].

Section 2 of this paper considers Canadian COVID-19 data and methods used to examine the spread of COVID-19 through various mathematical models including the SIR, SEIR and the SEIQRDP. It is to be noted that in the SIR and SEIR models, [R] represents removed cases (recovered + deceased), whereas in the SEIQRDP model, [R] represents only recovered cases. The graphs in this paper have been computed using Mathematica and Python to illustrate the transmission and spread trends of this disease over time. Section 3 discusses in detail the SEIQRDP model and its incorporation of protective measures. Various numerical methods have been applied in the context of this model including ordinary differential equations, fractional differential equations (using Matlab), and relating the generalised Lambert W function to this model. In section 4, the results of this paper are presented along with the numerical methods and analysis of the SIR, SEIR, SEIQRDP models using Canadian COVID-19 data. The final section comprises a summary of the findings and conclusions.

**Variants**

The majority of the mutations in RNA viruses, such as COVID-19, are expected to be deleterious. Generally, these mutations are not of concern, however, a fraction of the mutations can lead to variants impacting transmission rate, disease intensity, and vaccine efficacy. There are several variant classifications such as variants being monitored, variants of interest, variants of concern, and variants of high consequence. Variants of interest arise from a change in receptor binding that alters the transmissibility and severity of the disease. Along with this, variants classified in this category will be under increased surveillance in case the spread of this strain begins to dramatically increase. Variants of concern arise when there is an even higher risk of transmissibility and those infected with this strain are hospitalised. Along with treatments having a reduced effect, there is a rise in the number of deaths resulting from variants classified in this category. The alpha B.1.1.7, beta B.1.351, and gamma P.1 variants are all classified as variants of concern. Variants of high consequence currently have no strains classified under its banner. Strains under this classification have a significantly decreased response to treatment, causing scientists to eagerly work towards finding a treatment for the variant of high virulence. Vaccines have proven to be critical in the prevention of COVID-19 and while certain vaccines have shown some defence against certain variants, further studies are being done to generate an even more effective prevention tool. Mutations in viruses accumulate as time progresses, making it imperative to detect trends to help understand and prevent the spread of the variant [8].

**Methods**

In epidemiology, compartmental modelling is a generalized method to model infectious diseases. Using differential equations as a function of time, it is possible to take a closed population and map out the spread and transmission of disease as well as the interactions of its various compartments.

**SIR Model**

The SIR model, one of the simplest and oldest compartmental modelling techniques, consists of three components, Susceptible, Infected and Removed as functions of time in a closed population [10]. Susceptible individuals are those who are not infected but may become infected. The removed compartment includes individuals who were recovered or deceased. Recovered individuals can still be reinfected, returning them to the susceptible population [11], [12].

The normalized expressions for the SIR model are written below:

$$s(t) = \frac{S(t)}{N} \tag{1a}$$

$$i(t) = \frac{I(t)}{N} \tag{1b}$$

$$r(t) = \frac{R(t)}{N} \tag{1c}$$

When each compartment value is normalized by dividing the total population in the system, the three compartments should add up to 1. The sum of the normalized expressions equal to one:

$$s(t) + i(t) + r(t) = 1 \qquad (2)$$

Below are the differential equations for the SIR model that are used in disease tracking and progression:

$$\frac{dS}{dt} = -\beta s(t)i(t) \qquad (3a)$$

$$\frac{dI}{dt} = -\beta s(t)i(t) - \gamma i(t) \qquad (3b)$$

$$\frac{dR}{dt} = \gamma i(t) \qquad (3c)$$

where $\beta$ represents infection rate and $\gamma$ represents recovery rate [2], [13].

**SEIR Model**

The modified version of the SEIR model is an extension of the SIR model, with an added parameter "Exposed (E)" and time-dependent coefficients $\rho, q,$ and $r$. The SEIR model involves the same Susceptible-Infected-Removed population, but also considers the population that is incubating the virus, but is not infected or infectious, which is defined as Exposed [14]. The SEIR model equations can be seen below:

$$\frac{dS}{dt} = -\frac{\rho \beta SI}{N} \qquad (4a)$$

$$\frac{dE}{dt} = \frac{\rho \beta SI}{N} - q\alpha E \qquad (4b)$$

$$\frac{dI}{dt} = q\alpha E - r\gamma I \qquad (4c)$$

$$\frac{dR}{dt} = r\gamma I \qquad (4d)$$

where the parameters are defined as:

$\alpha$ = incubation rate from the exposed group to the infected group

$\beta$ = infection rate

$\gamma$ = recovery rate which is the transition from the infected group to the removed group

$\rho$ = time-dependent infection rate as a scaling coefficient

$q$ = incubation rate as a scaling coefficient, and

$r$ = recovery rate as a scaling coefficient.

These equations have been modified to fit a closed population where:

$$\frac{dS}{dt} + \frac{dE}{dt} + \frac{dI}{dt} + \frac{dR}{dt} = 0 \tag{5}$$

Time-dependent parameters such as $\beta$ and $\gamma$ help the model adapt to the combined effects of variants and protective measures. Scaling coefficients $\rho$, $q$, and $r$ are added to $\beta$, $\alpha$, and $\gamma$. When the SEIR model is applied to the modeling of the transmission of a particular variant, the three transmission parameters characteristic to the variant itself will be determined by the nature of the variant. The three scaling coefficients capture the effects of lockdowns, social distancing, quarantines, and PPE use. Ideally, by comparing the fitted value of those coefficients at different times, the result also gives us an idea of how effective the protective measures are [9].

**SEIQRDP Model**

The SEIQRDP model is an even further extension of the already extended SEIR model [15]. The SEIQRDP model introduces "Quarantining (Q)," "Death/Deceased (D)," and "Insusceptible (P)" parameters which provide a more accurate model of transmission[16]. The SEIQRDP equations can be seen below:

$$S'(t) = \frac{-\beta\big(S(t)I(t)\big)}{N} - \gamma S(t) \tag{6a}$$

$$E'(t) = \frac{\beta\big(S(t)I(t)\big)}{N} - \alpha E(t) \tag{6b}$$

$$I'(t) = \alpha E(t) - \delta I(t) \tag{6c}$$

$$Q'(t) = \delta I(t) - \lambda Q(t) - \kappa Q(t) \tag{6d}$$

$$R'(t) = \lambda Q(t) \tag{6e}$$

$$D'(t) = \kappa Q(t) \tag{6f}$$

$$P'(t) = \gamma S(t) \tag{6g}$$

along an independent variable, time, in a closed population, where *s(t), e(t), i(t), q(t), r(t), d(t),* and *p(t)* represent a numerical value over a selected population *N* and will add to 1:

$$s(t) + e(t) + i(t) + q(t) + r(t) + d(t) + p(t) = 1 \tag{7}$$

The parameters seen in the equations above are:

$\alpha$ = inverse of the average latent time

$\beta$ = infection rate

$\gamma$ = protection rate

$\delta$ = the rate at which infectious people enter quarantine

$\lambda$ = recovery rate, and

$\kappa$ = mortality rate.

A mathematical analysis of COVID-19 using the SEIQRDP model with a fractional differential approach was conducted, with a specific focus on the case studies of China, Algeria, Egypt, and Saudi Arabia [15]. The authors have given a rigorous mathematical analysis of the solutions. In our study we focus on the solutions they have presented and their analysis to the case study. Our focus is on the fractional differential equations (FDE) to use the fractional order models to provide a better fitting to real Canadian data. The use of FDE in mathematical modelling of biological phenomena has been used in the last few decades to better explain and process the various properties more accurately than integer order models [16].

Where $\lambda$ and $\kappa$ are time-dependent parameters of the logistic function given below, whereby we have chosen this form based on its various applications in disease modelling [15]. In the modified SEIQRDP model, we have introduced time-dependence in all parameters.

The modified SEIQRDP equations with time-dependent parameters can be seen below:

$$S'(t) = \frac{-\rho\beta\big(S(t)I(t)\big)}{N} - r\gamma S(t) \tag{8a}$$

$$E'(t) = \frac{\rho\beta\big(S(t)I(t)\big)}{N} - q\alpha E(t) \tag{8b}$$

$$I'(t) = q\alpha E(t) - a\delta I(t) \tag{8c}$$

$$Q'(t) = a\delta I(t) - b\lambda Q(t) - c\kappa Q(t) \tag{8d}$$

$$R'(t) = b\lambda Q(t) \tag{8e}$$

$$D'(t) = c\kappa Q(t) \tag{8f}$$

$$P'(t) = r\gamma S(t) \tag{8g}$$

The equations for the time-dependent parameters can be seen below:

$$a = b = \rho = q = \frac{1}{1 + e^{-t}} \tag{9a}$$

$$c = \frac{1}{e^{-t} + e^{t}} = \frac{1}{2}\operatorname{sech}(t) \tag{9b}$$

$$r = \frac{0.8}{1 + e^{-t}} \tag{9c}$$

9(a) and (c) are logistic sigmoid functions and (9b) is a half-hyperbolic secant function, which has many applications in statistics, biochemistry and pharmacology. This allows us to better use data to model Covid-19.

Fractional differential equations were also used in the original SEIQRDP model. They can be seen below:

$$^{C}D_0^a S(t) = -\beta \frac{S(t)I(t)}{N_{pop}} - \gamma S(t) \tag{10a}$$

$$^{C}D_0^a E(t) = \beta \frac{S(t)I(t)}{N_{pop}} - \alpha E(t) \tag{10b}$$

$$^{C}D_0^a I(t) = \alpha E(t) - \delta I(t) \tag{10c}$$

$$^{C}D_0^a Q(t) = \delta I(t) - \lambda(t)Q(t) - \kappa(t)Q(t) \tag{10d}$$

$$^{C}D_0^a R(t) = \lambda(t)Q(t) \tag{10e}$$

$$^{C}D_0^a D(t) = \kappa(t)Q(t) \tag{10f}$$

$$^{C}D_0^a P(t) = \gamma S(t) \tag{10g}$$

Where $\lambda(t)$ and $\kappa(t)$ are defined below:

$$\lambda(t) \in \left\{ \frac{\lambda_0}{1 + \exp(-\lambda_1(t - \lambda_2))}, \quad \lambda_0 + \exp(-\lambda_1(t + \lambda_2)) \right\} \tag{11}$$

$$\kappa(t) \in \left\{ \frac{\kappa_0}{\exp(-\kappa_1(t - \kappa_2)) + \exp(\kappa_1(t - \kappa_2))}, \kappa_0 \exp(-[\kappa_1(t - \kappa_2)]^2), \kappa_0 + \exp(-\kappa_1(t + \kappa_2)) \right\} \tag{12}$$

Empirical coefficients are needed to tune the time-dependency of the parameters.

The solutions of these fractional differential equations can be generalized as:

$$a(t) = a_n + \frac{1}{G(a)} \int_0^t (t - s)^{a-1} f_n(s, X(s)) ds \tag{13}$$

e.g,
$$S(t) = S_0 + \frac{1}{G(a)} \int_0^t (t - s)^{a-1} f_1(s, X(s)) ds$$

We used Runge-Kutta 4th order numerical method in our manuscript [17]. The original Runge-Kutta 4th order method can be seen below:

$$K_1 = h f(x_n, y_n) \tag{14a}$$

$$K_2 = hf\left(x_n + \frac{h}{2}, y_n + \frac{k_1}{2}\right) \tag{14b}$$

$$K_3 = hf\left(x_n + \frac{h}{2}, y_n + \frac{k_2}{2}\right) \tag{14c}$$

$$K_4 = hf(x_n + h, y_n + k_3) \tag{14d}$$

$$y_{n+1} = y_n + \frac{k_1}{6} + \frac{k_2}{3} + \frac{k_3}{6} + \frac{k_4}{6} + O(h^5) \tag{14e}$$

The fractional Runge-Kutta 4th order method can be seen below [17]:

$$Y_{i+1} = Y_i + \frac{1}{6}(K_1 + 2K_2 + 2K_3 + K_4) \text{ with } \kappa = \frac{h^a}{\Gamma(a)} \text{ and} \\ K_1 = \kappa f(X_i, Y_i), \tag{15a}$$

$$K_2 = \kappa f\left(X_i + \frac{1}{2}\kappa, Y_i + \frac{1}{2}K_1\right) \tag{15b}$$

$$K_3 = \kappa f\left(X_i + \frac{1}{2}\kappa, Y_i + \frac{1}{2}K_2\right) \tag{15c}$$

$$K_4 = \kappa f(X_i + \kappa, Y_i + K_3) \tag{15d}$$

In the SEIQRDP model with fractional equations, the Caputo fractional derivative of order a is used instead of the standard derivative. The Caputo fractional derivative is a way to generalize the concept of a derivative to non-integer orders, and it has been found to be useful in modeling complex systems with memory effects.

In this case, the SEIQRDP model with fractional equations allows for a more accurate representation of the spread of COVID-19 by accounting for the long-term memory effects of the disease. This approach considers the fact that infected individuals may take a longer time to recover or may continue to spread the disease even after recovering.

The parameters in the fractional SEIQRDP equations have the same interpretation as in the standard SEIQRDP model, and they represent the various rates of disease transmission, recovery, and mortality. The use of fractional calculus in this model provides a more realistic and accurate representation of the spread of COVID-19 and can aid in the development of effective control and prevention strategies.

**Results**

The figures below display the fitted fractional SEIQRDP model results and their corresponding residual plots using a Canadian COVID-19 dataset. The results generated using Matlab (version R2022b) focus on the number of cases quarantined (Q), recovered (R), and death/deceased (D). The fitted parameters for the plots were a = 0.94, 0.96, 0.98, 1, 1.02, 1.06, and 1.1. a = 1 represents the Ordinary Differential Equation (ODE) while all other "a" values represent Fractional Differential Equations (FDE). Plots with a smaller residual error indicate a better fit with the COVID-19 data as demonstrated by a smaller area under the curve. The authors focused on two waves of COVID-19 in Canada with the first wave spanning days 50 to 100 (20/04/2020 – 03/06/2020) and the second wave spanning days 200 to 320 (11/09/2020 – 09/01/2021). The dotted line corresponds to the Canadian data and the solid lines refer to predicted trends in the fractional plots [18].

Quarantined

Quarantined data has been analysed according to each COVID-19 wave, as outlined above. When fitting "a" values less than 1 (Figure 1a), a general fit was best suited to ODE, however an a value of 0.98 (yellow) was favoured on June 3. Overall, Figure 1a showed a lot of variability among data points with transitions across all the FDE a-values. The corresponding residual plot (Figure 1b) showed a fit favouring a = 0.94 (purple) from 27/04/2020 until 09/05/2020. From 11/05/2020 - 22/05/2020 ODE was best suited to the data points. The end of Figure 1a best fit a = 0.94 (purple). When fitting "a" values greater than 1 (Figure 7a), a = 1.1 (red) was favoured from 20/04/2020 to 02/05/2020. The remaining data points transitioned between the ODE model and a = 1.1 (red). The corresponding residual plot (Figure 7b) had an overall uniform appearance with no dominant parameter.

When fitting "a" values less than 1 for the second wave (Figure 2a), ODE showed the best fit for the data at the beginning and end of the wave. FDE parameter a = 0.94 (green) showed the best fit from the second week of November 2020 through the second week of December 2020. The corresponding residual plot (Figure 2b) showed a = 0.96 (yellow) as the best fit from the beginning of the second wave until the last week of September 2020. Several transitions between a = 0.98 (red) and a = 0.94 (purple) were seen until the second week of November 2020. FDE parameter a = 0.96 (yellow) was favoured from the second week of November 2020 to the first week of December 2020. The remaining data points in the second wave favoured a = 0.94 (purple). When fitting "a" values greater than 1 (Figure 8a), a = 1.1 (red) best suited the data from the beginning of the second wave until the second week of November 2020. The remaining data points favoured the ODE model. The corresponding residual plot (Figure 8b) favoured a = 1.1 (blue) from the beginning of the second wave until the fourth week of October 2020 and the second week of November 2020 until the second week of December 2020. The remaining data points alternate between a = 1.06 (red) and the ODE model.

Recovered

Recovered data has been analyzed according to each COVID-19 wave, as outlined above. When fitting "a" values less than 1 (Figure 3a), a fit best suited to ODE was observed at most data points. However, from 07/05/2020 - 24/05/2020, a = 0.98 (yellow) was observed to be the best fit. The corresponding residual plot (Figure 3b) shows a = 0.94 (purple) as the best fit from 06/05/2020 - 11/05/2020. The remaining data points have an overall uniform appearance with no dominant parameter. When fitting "a" values greater than 1 (Figure 9a), a = 1.1 (orange) best fits the data from 20/04/2020 - 30/04/2020. The remaining data points in the first wave were best suited to the ODE model. The corresponding residual plot (Figure 9b) displayed a = 1.1 (blue) as the best fit from 24/04/2020 - 05/05/2020. The remaining data points have an overall uniform appearance with no dominant parameter.

When fitting "a" values less than 1 for the second wave (Figure 4a), ODE showed to be the best fit for data points after the first week of November 2020. Prior to November 2020, FDE parameters showed to be the best fit. The corresponding residual plot (Figure 4b) showed a =

0.96 (yellow) as the best fit from the beginning of the second wave to the start of October 2020. The remaining data points showed a fit favouring the FDE parameters, with a = 0.94 showing the overall best fit, however this is inconsistent with the data in Figure 4a. When fitting "a" values greater than 1 (Figure 10a), ODE showed to be the favourable fit for the data points, however a = 1.1 (orange) showed to be the best suited parameter until the second week of October 2020. The corresponding residual plot (Figure 10b) is consistent with the data shown in Figure 10a. The FDE parameter a = 1.1 is best suited until the second week of October 2020, with the remaining data favouring the ODE model.

<u>Death/Deceased</u>

Death/deceased data has been analysed according to each COVID-19 wave, as outlined above. When fitting "a" values less than 1 (Figure 5a), a fit best suited to the ODE model was observed with brief transitions of a = 0.98 (yellow). The corresponding residual plot (Figure 5b) did not have a dominant parameter. FDE parameter a = 0.94 (purple) best suited the data from 06/05/2020 to 17/05/2020, however great amounts of variability between parameters was seen for the remaining first wave data points. When fitting "a" values greater than 1 (Figure 11a), a fit best suited to the ODE model was observed with a = 1.02 (purple) temporarily dominating from 18/05/2020 to 26/05/2020. The corresponding residual plot (Figure 11b) had no dominant "a" value for the first wave data points.

When fitting "a" values less than 1 (Figure 6a), a fit best suited to ODE was observed at most data points. FDE parameters did not show a strong fit to the second wave data points. Conversely, the corresponding residual plot (Figure 6b) did not show pure dominance for ODE. The ODE model was best suited from the third week of October to the third week of November. However, the remaining data points showed favourable fits to several FDE parameters. When fitting "a" values greater than 1 (Figure 12a), a fit best suited to ODE was observed at most data points. The corresponding residual plot (Figure 12b) was best suited to a = 1.02 (yellow) for the majority of the data points. However, small gaps were seen where the remaining parameters were best suited for the data.

**Discussion**

Model Interpretations

In this study, the practice of social distancing and self-quarantine was considered as mechanisms to reduce disease transmission and spread [19]. Placing an individual into quarantine is the act of establishing a strict isolation to prevent the spread of disease (1). It is important to note that the quarantined population only includes infected individuals. It does not include individuals who were in quarantine, but not infected. When increasing the number of compartments from the SIR to the SEIQRDP model the accuracy of the model improves, indicated that additional compartments account for more features.

This paper extends on our previous paper "A mathematical model of COVID-19 transmission" (2022) [9] where the focus was on modeling covid data using minimal parameters and the basic variations of the SIR model. In this paper, we aimed to add additional parameters, accounting for the use of non-pharmaceutical interventions including the use of PPE, social distancing and lockdown measures. This incorporates the behavioural changes resulting from the use of such interventions and thus allowing the assessment of its impact on the spread of COVID-19. These were accounted for in our models through the use ODEs and FDEs to create more accurate plots which better reflect the Canadian population. By using these plots–we can extrapolate Canadian behavior and reaction to a health crisis which can allow us to evaluate our health/crisis plans [18], [20].

Unlike ODE's, FDE's can measure non-local effects which allows for an analysis across a broader time frame. Thus, non-pharmaceutical interventions such as quarantining, and vaccination can be modelled and analyzed using FDE's. FDE's can account for more anthropological factors compared to the ODE model, such as infectious asymptomatic carriers. FDE's allow for a greater manipulation of time dependent equations, while they do present a difficult method of model analysis. This was seen through the degree of interpretation of the graphs which can be increased through the acquisition of a larger data sample size. Conversely, the ODE model generally provided better fits for the number of recovered and deceased cases, indicating more regular patterns in these variables[21].

This analysis demonstrates that both the ODE and FDE models can be effective in capturing different aspects of the COVID-19 dynamics. The choice of the model and its parameter values depends on the specific variable being analyzed and the stage of the pandemic. Through the incorporation of additional parameters one can capture the effectiveness of various non-pharmaceutical interventions to prevent and reduce the spread and probability of transmission. By quantifying the reduction in transmission risk associated with these interventions, our models aim to estimate the overall impact on the spread of the virus within the population[18], [20].

Compartmental modeling has been shown to accurately represent COVID-19 transmission. In this paper, we increased the variables from the classic SIR model to the SEIQRDP model to test whether this improves the accuracy from a Canadian perspective [cite paper regarding SEIQRDP model]. We find that increasing the number of variables/compartments and generalizing to FDEs, leads to a better representation of the realistic features of the disease dynamics for the COVID-19 mathematical model. We fitted the parameters to different orders of the fractional differential equations and found that for values of a greater than 1, better fits were obtained. It was important to have different values of parameters for different orders of the fractional differential equations to have correct fits. The use of fractional differential equations was shown to also work effectively when analyzing Algerian data[15].

It is important to note that the results presented in this study are based on a specific Canadian COVID-19 dataset and the fitted parameters obtained using Mathematica. Therefore, caution should be exercised in generalizing these findings to other regions or populations. Future research could explore the applicability of the SEIQRDP model with fractional derivatives in different contexts and compare its performance with other epidemic models to gain further insights into the dynamics of infectious diseases like COVID-19.

These models provide insight on crisis responses of the Canadian population while illustrating the deficiencies of the Canadian health system and how they can be improved. The eventual shortage of PPE availability in Canada demonstrated its criticality during the COVID-19 pandemic [22]. This presents an opportunity for possible improvements in infrastructure to better prepare future crisis plans and emergency protocols.

## Conclusions

The emergence of computers, artificial intelligence, and remote work has assisted in the moderate control of the pandemic. More effective, ethical, and wise use of machine learning and artificial intelligence should be a priority for better management and control of the spread of future epidemics and pandemics [6], [23], [24]. COVID-19 has led to chronic ailments such as long COVID, a complex and debilitating condition. It has more than 50 identified symptoms of which fatigue, respiratory problems, and cognitive impairments are dominant. Management of long COVID is an important priority as the effects of COVID-19 are longstanding. According to Health Canada, approximately 1.3 million Canadians are suffering from long COVID.

Previous pandemics such as SARS and the Spanish flu have vastly impacted the course of human history, leaving behind profound societal changes [14], [19]. In terms of COVID-19, the effects of rapid 21st century globalization led to the rapid spread of the virus which was contrary to what was seen in the past with SARS. It exposed vulnerabilities in healthcare systems and economic structures. However, advancements in communication, technology, and science, specifically the rapid development of vaccines equipped us with tools to combat the virus more effectively than before. Valuable lessons from previous pandemics undoubtedly played a crucial role in shaping our response to COVID-19, underscoring the importance of preparedness, global cooperation, and equitable access to healthcare resources. Due to political and corporate interests, vaccines and their patents were not shared with developing countries, which led to global vaccine inequity [6], [20]. This, in turn led to longer times for COVID-19 virus to further mutate and wreak higher losses especially to poor and vulnerable populations. A key lesson, therefore, is that equitable and free distribution of vaccine patents is not only ethical but also critical for minimizing losses from upcoming pandemics.

Fractional order differential equations have been successfully used in applications in materials engineering, physical, and biological sciences [25]. In this paper we have worked on the numerical fits of the SEIQRDP model using both integer and fractional order differential equations. We have shown that in some cases with fractional order (a>1) better fits were obtained for the data. We used the fractional Runge-Kutta method [15], [17]. The data was obtained from a Canadian COVID-19 dataset. Fractional order differential equations might be more appropriate than standard differential equations due to better fits of the data when there is greater variability. More work on the fractional differential equations approach using the Laplace Transform approach is currently in progress [18].

One must accept the reality of the continuing existence of COVID-19 and other deadly viruses and take appropriate protective measures to protect themselves and others. Protective measures must be implemented and enforced to ensure that such diseases do not cause such catastrophic implications to the global population.

## Acknowledgements

We thank Prof. Jaclyn Duffin at the Departments of Philosophy, History at Queens University for referring us to her book "COVID-19: A History"[22].

**Supplemental Figures**

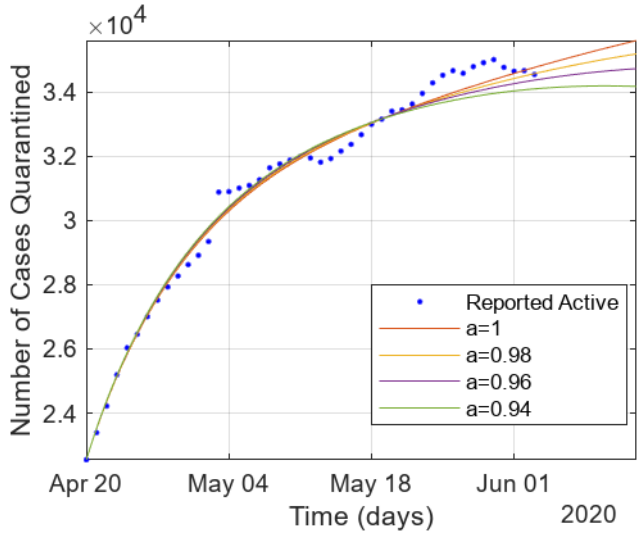

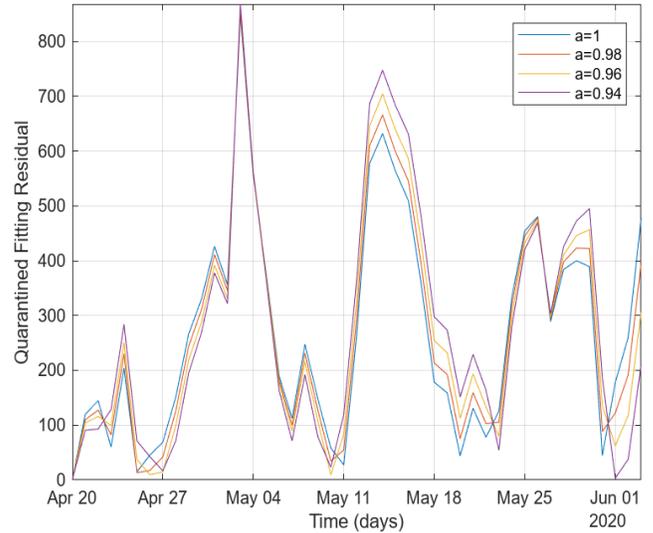

**Figure 1a.** Number of quarantined individuals fitted was a-values less than 1 in COVID-19 first wave

**Figure 1b.** Corresponding residual plot of Figure 1a fitted with a-values less than 1

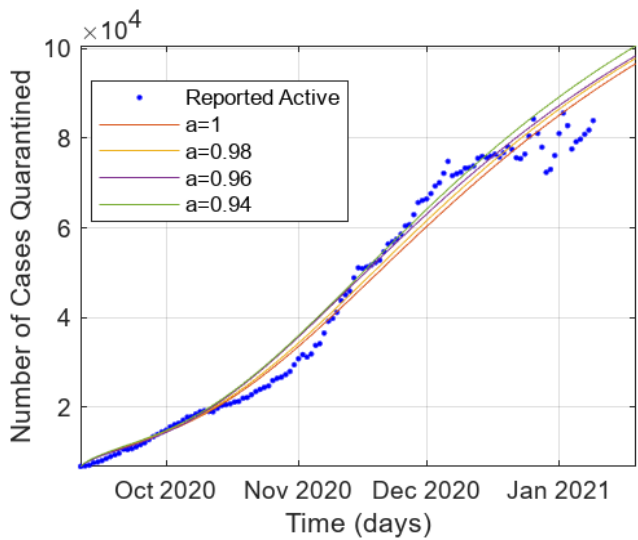

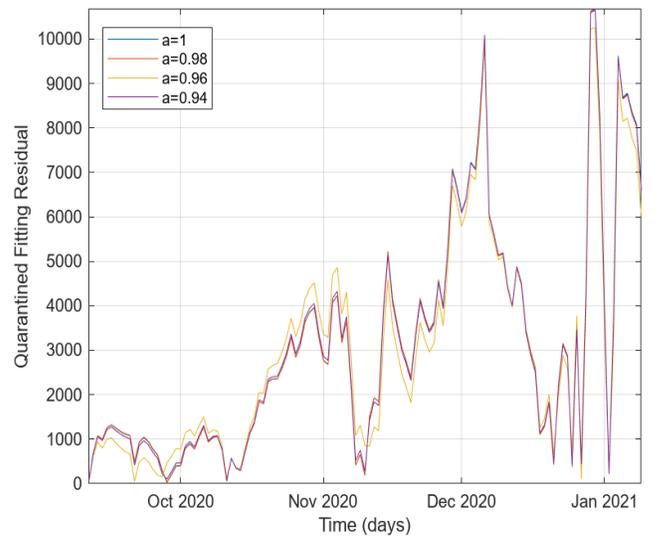

**Figure 2a.** Number of quarantined individuals fitted was a-values less than 1 in COVID-19 second wave

**Figure 2b.** Corresponding residual plot of Figure 2a fitted with a-values less than 1

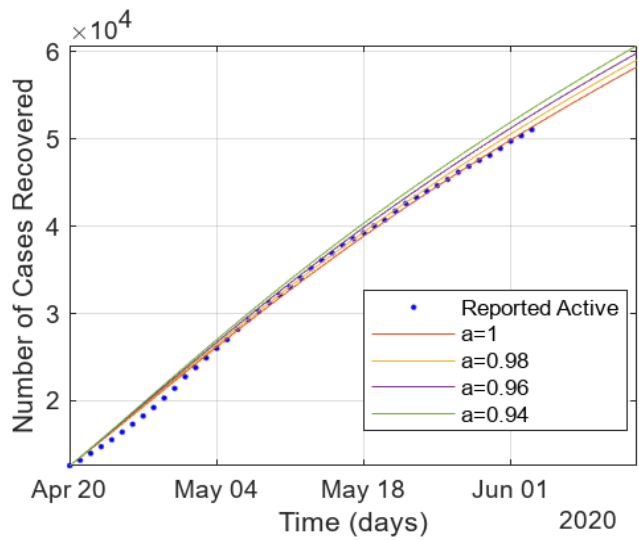

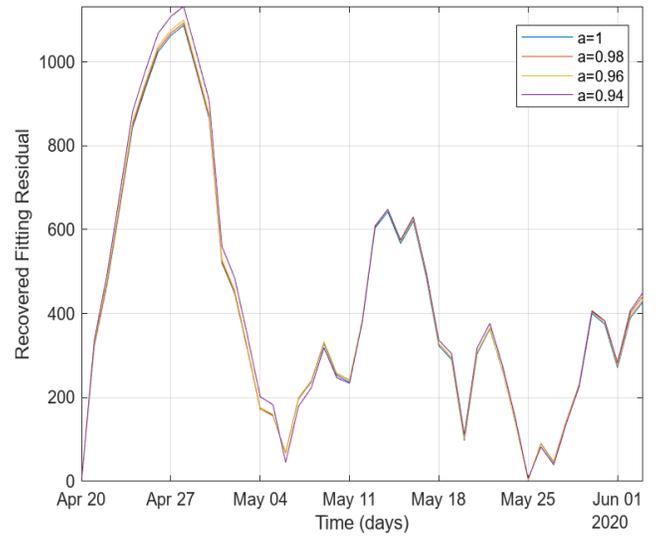

**Figure 3a.** Number of recovered individuals fitted was a-values less than 1 in COVID-19 first wave

**Figure 3b.** Corresponding residual plot of Figure 3a fitted with a-values less than 1

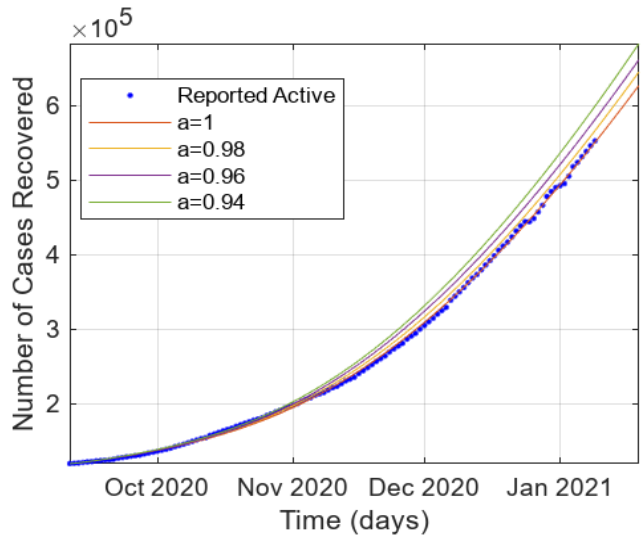

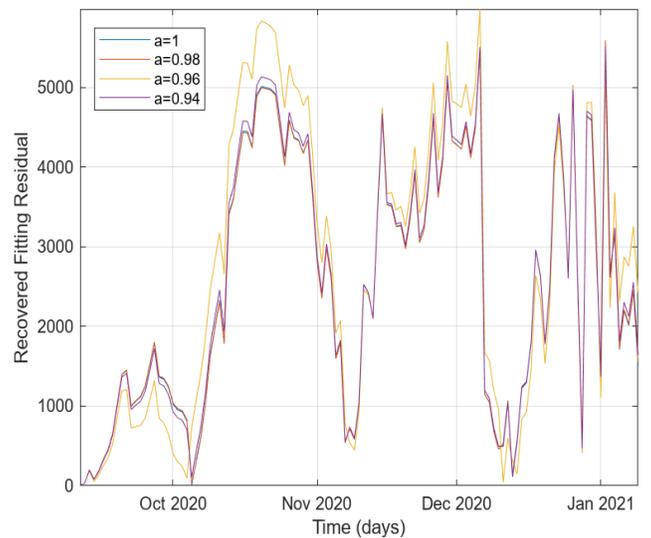

**Figure 4a.** Number of recovered individuals fitted was a-values less than 1 in COVID-19 second wave

**Figure 4b.** Corresponding residual plot of Figure 4a fitted with a-values less than 1

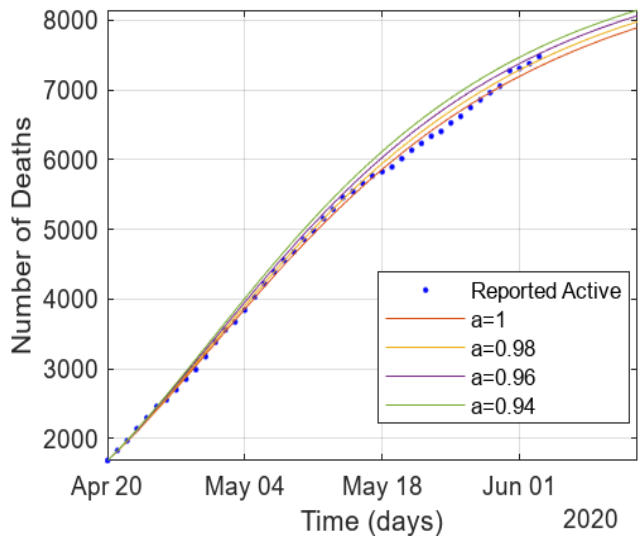

**Figure 5a.** Number of deceased individuals fitted was a-values less than 1 in COVID-19 first wave

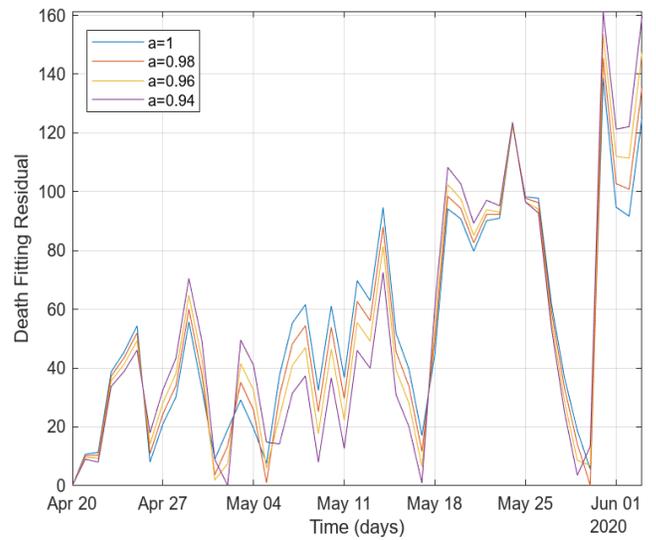

**Figure 5b.** Corresponding residual plot of Figure 5a fitted with a-values less than 1

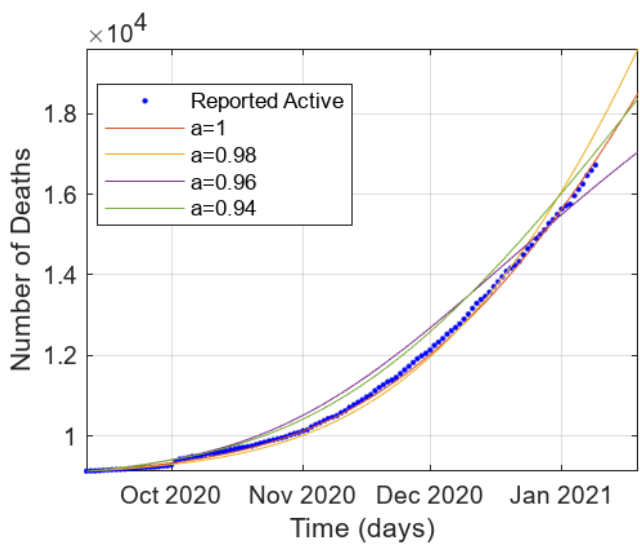

**Figure 6a.** Number of deceased individuals fitted was a-values less than 1 in COVID-19 second wave

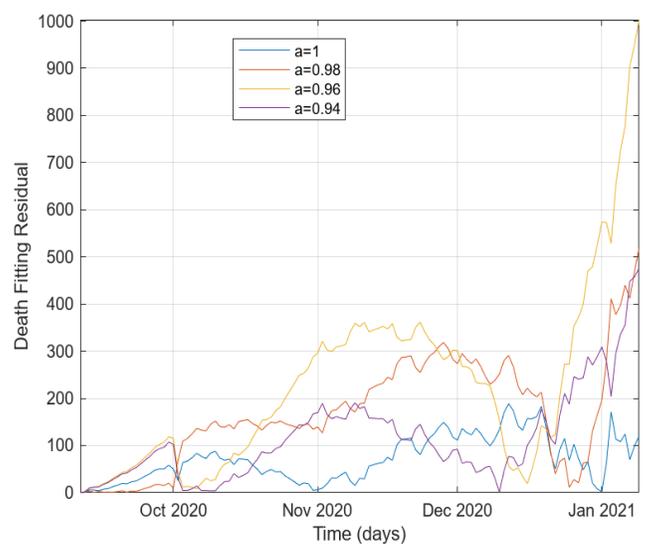

**Figure 6b.** Corresponding residual plot of Figure 6a fitted with a-values less than 1

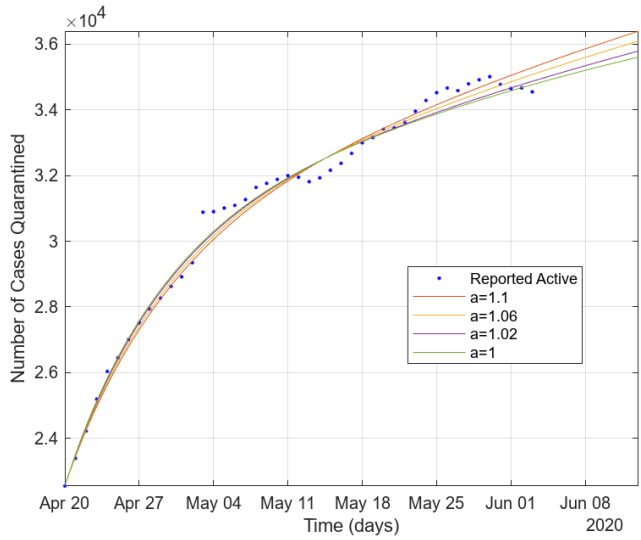

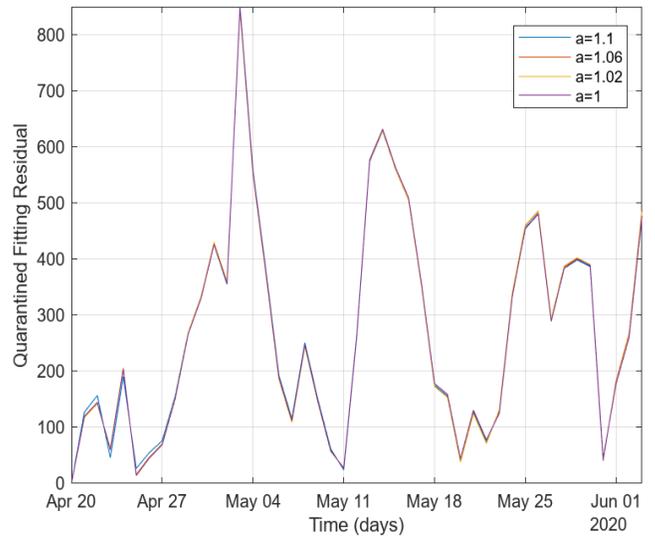

**Figure 7a.** Number of quarantined individuals fitted was a-values greater than 1 in COVID-19 first wave

**Figure 7b.** Corresponding residual plot of Figure 7a fitted with a-values greater than 1

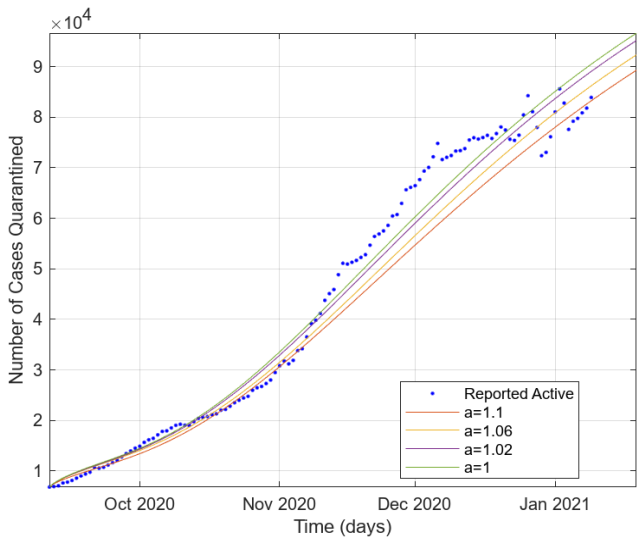

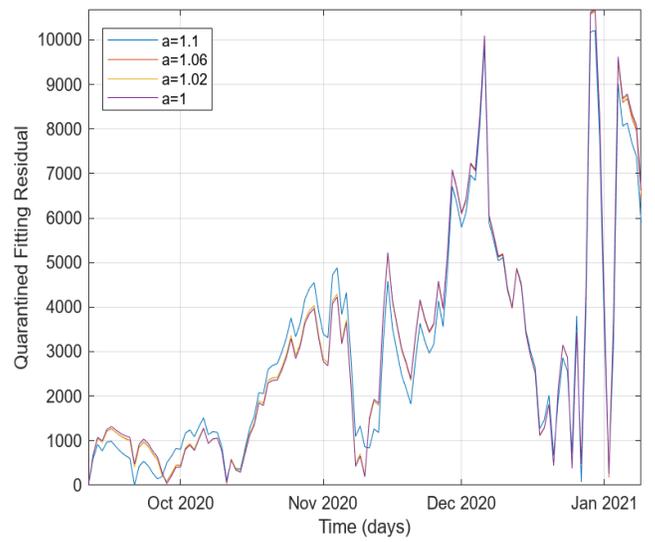

**Figure 8a.** Number of quarantined individuals fitted was a-values greater than 1 in COVID-19 second wave

**Figure 8b.** Corresponding residual plot of Figure 8a fitted with a-values greater than 1

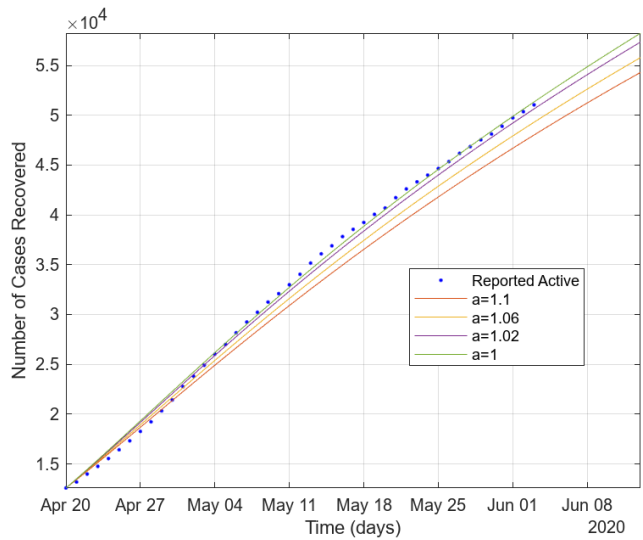

**Figure 9a.** Number of recovered individuals fitted was a-values greater than 1 in COVID-19 first wave

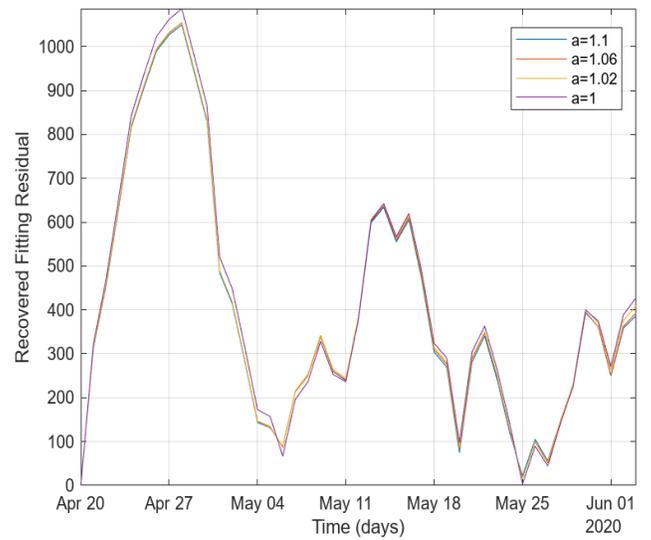

**Figure 9b.** Corresponding residual plot of Figure 9a fitted with a-values greater than 1

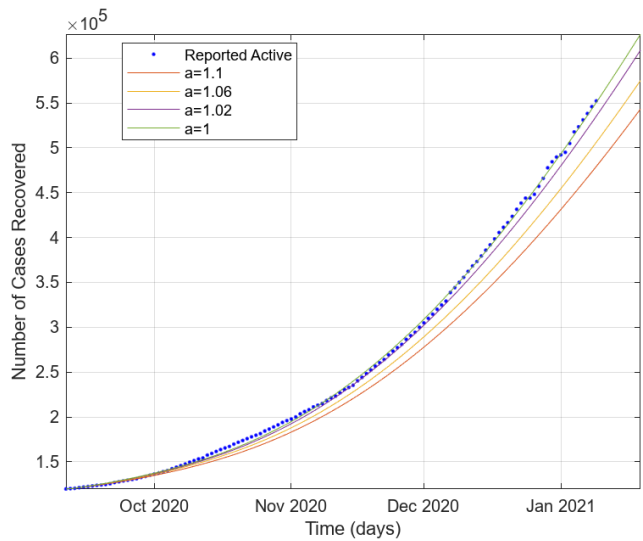

**Figure 10a.** Number of recovered individuals fitted was a-values greater than 1 in COVID-19 second wave

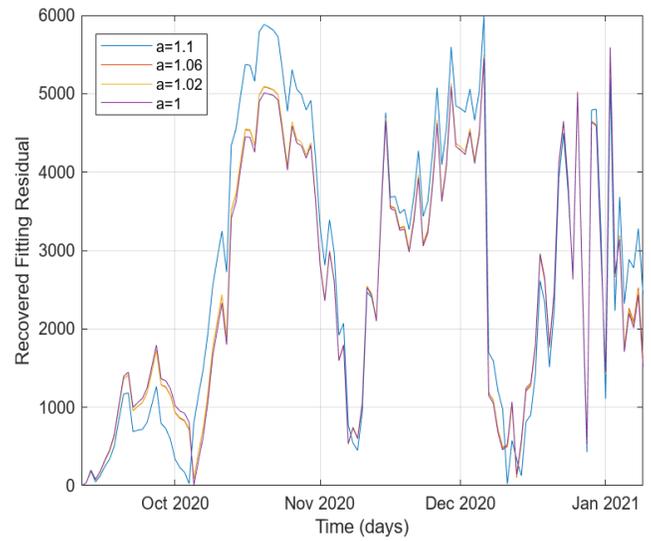

**Figure 10b.** Corresponding residual plot of Figure 10a fitted with a-values greater than 1

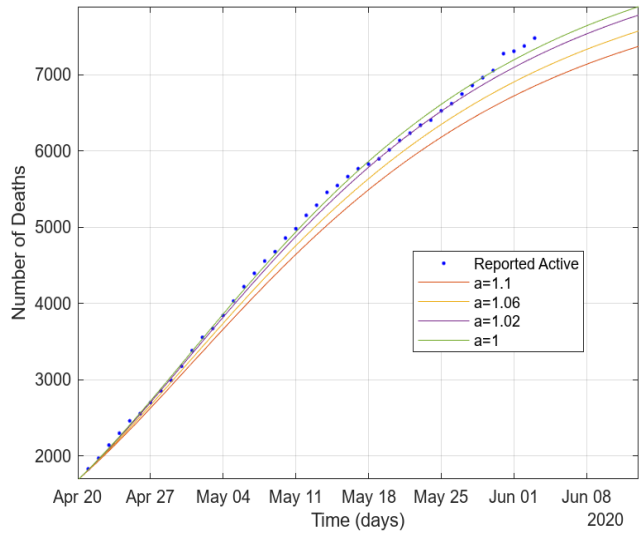

**Figure 11a.** Number of deceased individuals fitted was a-values greater than 1 in COVID-19 first wave

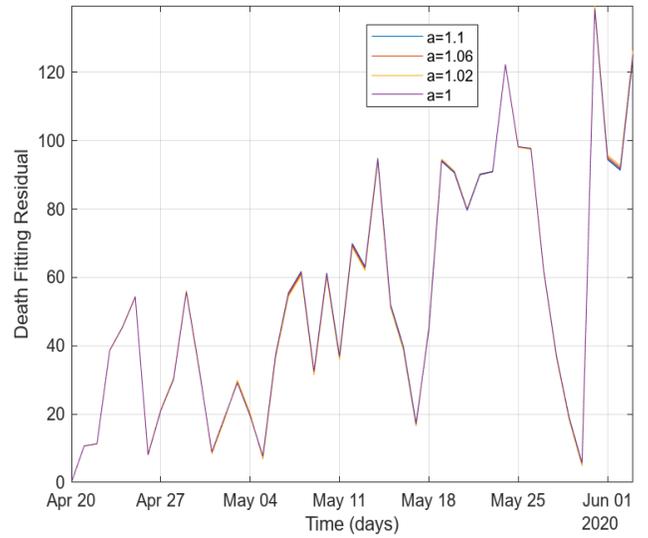

**Figure 11b.** Corresponding residual plot of Figure 11a fitted with a-values greater than 1

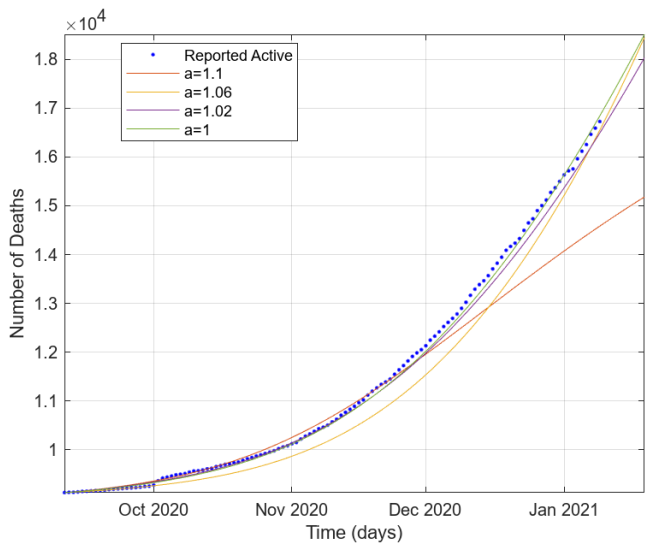

**Figure 12a.** Number of deceased individuals fitted was a-values greater than 1 in COVID-19 second wave

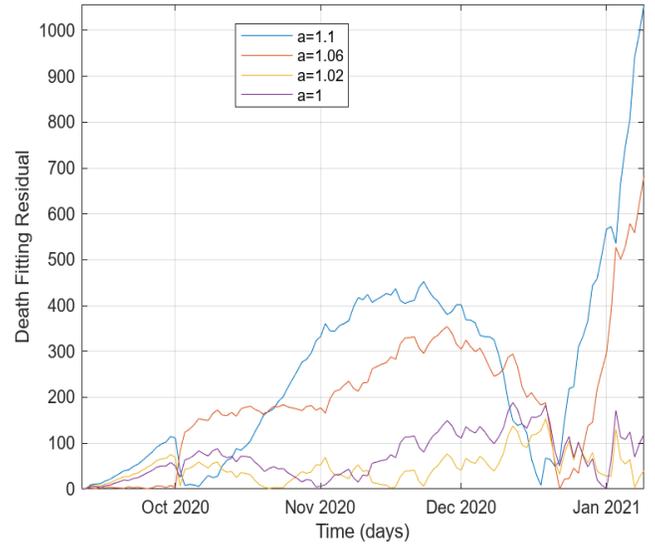

**Figure 12b.** Corresponding residual plot of Figure 12a fitted with a-values greater than 1